# A Centralized Channel Allocation Method in Clustered Ad Hoc Networks
## (Imperialist Competitive Algorithm)

Mahboobeh Parsapoor , Urban Bilstrup

*Abstract*— Cognitive radio networks (CRNs) is the next generation of wireless communication. This type of network requires efficent spectrum allocation methods. This paper presents a new meta-heuristic evolutionary method for solving the channel allocation problem in an ad hoc network context. The suggested method is based on a graph-theoretic model and seeks a solution for the spectrum allocation problem in a clustered ad hoc network topology.The method is referred to as imperialist competitive algorithm (ICA)and provides a scheme for allocating the available channels to cluster heads maximizing spectrum efficiency and minimizes co-channel interference. The suggested methods are tested for several scenarios; the performance of the ICA-based scheme is compared with the genetic algorithm based scheme.

## I. INTRODUCTION

The focus of this paper is on developing a new combinatorial optimization method for wireless ad hoc network that is the foundation of the cognitive ad hoc network. This type of network will require a large bandwidth; which is often a major issue; referring to spectrum scarcity. In order to maximize spectral utilization and overcome the spectrum limitation, spatial channel reuse schemes are needed.

In wireless communication, finding an effective channel allocation scheme is referred to as the channel allocation problem has been identified as an NP-hard class of problem [1]. An high utilization of the available channels means minimization the number of used channels while satisfying the interference constraints [1]. However, an efficient scheme for allocating a minimum number of channels, (chromatic index) to the wireless nodes satisfying the proposed constraints would not be found in polynomial time. However, heuristic methods, such as genetic algorithms (GA), swarm intelligence (SI) and ant colony optimization (ACO), have the capability to find near optimal solutions in polynomial time [1]-[4].

This paper suggests the use of a new evolutionary algorithm for solving the channel allocation problem in clustered ad hoc networks. This method is known as imperialist competitive algorithm (ICA) and is a graph theoretic-based method. The considered optimization criteria are: maximizing spectral efficiency and minimizing the co-channel interference between clusters.

The rest of this paper is formed as follows: Section II gives a brief review of the related works in spectrum allocation. In Section III, the system model is presented and the problem is formulated. Section IV illustrates ICA and the ICA-based spectrum allocation scheme. In Section V, we present the results of applying the ICA-based algorithm to several scenarios. We also compare the perfmance of the ICA-based scheme with the GA-based scheme. In Section VI, we conclude with some final notes.

## II. RELATED WORKS IN SPECTRUM ALLOCATION

As was mentioned, one of the challenging tasks in wireless communication is finding an efficient scheme for channel allocation. In wireless commination literature, the spectrum allocation has been mentioned as an optimization problem and numerous optimization algorithms have been proposed for that [1]-[4]. These algorithms can be categorized into several types of approach: graph-theoretic,, game-theoretic  and machine learning-based approaches. From the perspective of graph-theoretic approaches, the channel allocation problem is equivalent to the graph coloring problem, which is a well-known problem in graph theory.  Numerous graph-theoretic based optimization methods have been proposed for channel allocation schemes in cellular networks, ad hoc networks and cognitive radio networks [1]-[13]. One main issue of the suggested channel allocation schemes is the hidden terminal problem. Hence, the cluster-based spectrum allocation methods have been developed to allocate channel to the radio nodes avoiding this problem [5]. Heuristic method (e.g., greedy) and meta-heuristic methods (e.g., genetic algorithms (GAs) and ant colony optimization (ACO)) have been applied to solve this problem, which earlier has been referred to as a 'cluster based coloring algorithm' [5]. In the following we mention some studies that have investigated the channel allocation problem. A distributed cluster-based channel allocation scheme as a 'color-based clustering algorithm' has been suggested in [5]. In [6], a TDMA cluster-based multi-channel algorithm has been introduced for both inter-clustering and intra-clustering scheduling. The model aims to maximize the throughput while minimizing the number of allocated time slots avoiding co-channel interference between the cluster heads.
Recently, graph-theoretic based schemes for solving spectrum allocation problem in CRNs have been proposed. As example, a centralized spectrum allocation scheme has been applied to assign unused channels in the TV white space bands to IEEE 802.22 base station [7]. In order to maximize spectral efficiency, both centralized and distributed channel allocation protocol has been proposed by [8]. A centralized channel allocation method with the ability to transport the desired traffic, avoiding interference, is given in [9]. A distributed channel allocation method has been introduced to minimize interference and implementation complexity [10].

### III. RELATED ASSUMPTION

This paper investigates channel allocation problem in wireless ad hoc networks with the hybrid infrastructure that is similar to Centralized Ad hoc Network Architecture (CANA) [11]. It consists of a central access point and the wireless nodes that uses a distributed clustering algorithm (e.g., Lowest ID), the wireless nodes form the clustered topology. A common clustered network topology encompasses three types of mobile nodes that have been categorized as *cluster head, gateway* and *ordinary nodes* (see Figure 1). The cluster head, the master of a cluster, is responsible for resources scheduling and coordinates the intra cluster communication. The gateway, which is a common node between two or more clusters, provides the connectivity between the clusters. Others nodes are ordinary nodes that determine the boundary of clusters [12]-[14]. After forming clusters, only cluster heads are connected to the access point and a centralized channel allocation algorithm can be applied by the central access point.

The main assumptions for simulating this network are summarized as follows: 1) we simulate a snapshot of the network where the centralized controller executes the channel allocation algorithm to assign the available channels (code, time or frequency) to the received demands. 2) The allocated channels are orthogonal channels that can be exclusively used by each cluster head for intra cluster scheduling. 3) We have omitted to explain how an inter cluster communication can be managed and only considered the channel allocation to the demands for intra cluster communication. 4) No models for transmission activity and nodes' mobility have been considered. 5) Each node has an omni-directional antenna. 6) All of the nodes use similar transmission power, which will be unchanged during the channel allocation procedure. 7) To maximize the spectrum utilization, the same channels can be assigned to the cluster heads that are sufficiently far from each other (i.e., spatial channel reuse).

### IV. SYSTEM MODEL AND PROBLEM FORMULATION

The clustered ad hoc network can be represented using an undirected graph $G(V,E)$. The cluster heads are associated to the vertices, $V$ and $E$ is the set of edges; each edge shows the mutual neighbor relationship between cluster heads (see Fig. 1).

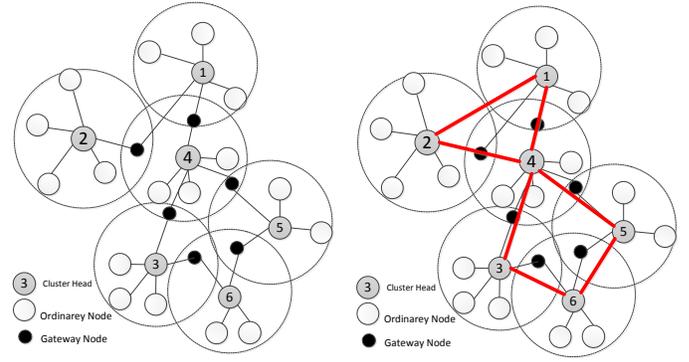

Figure 1. A clustered ad hoc network; the $G(V, E)$ is determined by the graph.

The clustered ad hoc network has a set of cluster heads as $\{C_1, C_2, C_3, ..., C_{N_{cluster}}\}$ and a set of available channels as $\{ch_1, ..., ch_{N_{Avaiable\_ch}}\}$ where $N_{cluster}$ and $N_{Avaiable\_ch}$ are the number of clusters and the number of available channels, respectively. A channel allocation scheme can be determined as an $N_{Avaiable\_ch} \times N_c$ matrix, $\mathbf{x}$; where $x_{pq} = 0$ or 1. If $p^{th}$ channel, $ch_p$, is assigned to $q^{th}$ cluster, $C_q$, then $x_{pq} = 1$ otherwise it is 0. The optimization functions can be formulated according to equation (1); an optimal solution minimizes the number of used channels. We also define an $N_c \times N_c$ matrix as $\mathbf{y}$; where $y_{pq} = 0$ or 1. If $p^{th}$ cluster, $C_p$ is one of neighbors of $q^{th}$ cluster, $C_q$, then $x_{pq} = 1$ otherwise it is equal to zero. Equation (2) formulates the constraint that is defined to avoid assigning the same channels to the neighbor cluster heads. Here $\mathbf{z}$ is the result of multiplication operation between two matrices $\mathbf{y}$ and $\mathbf{x}$.

$$\min \sum_{p=1}^{N_{Avaiable\_ch}} \sum_{q=1}^{N_c} x_{pq} \qquad (1)$$

$$s.t. \sum_{p=1}^{N_c} \sum_{\substack{q=1 \\ p \neq q}}^{N_c} z_{pq} = N_c \qquad (2)$$

$$\mathbf{z} = \mathbf{y} \times \mathbf{x} \qquad (3)$$

### V. ICA-BASED SPECTRUM ALLOCATION SCHEME

As was mentioned, we suggest Imperialist Competitive Algorithm (ICA) to solve channel allocation problem. In next subsection, we explain how ICA performs to find an optimal solution.

#### A. Imperialist Competitive Algorithm (ICA)

ICA is a new meta-heuristic optimization method inspired by "*imperialist competition*"[16]. In ICA, each individual is named '*country*' [25] as (4). As a matter of fact, a vector of optimization parameters $a_i$ is named as '*country*' [16].

$$country = \overline{a_1,...,a_n} \quad (4)$$

The steps of this algorithm can be summarized as follows:
1) An initial population that is a set of countries with different values for the optimization parameters is generated.
2) The power of each country is calculated according to: (5). Here $Cost_i$ the cost function of $i^{th}$ country is defined on the basis of the objective function.

$$p_i = \left| \frac{Normalized\_Cost_i}{\sum_{i=1}^{n_{pop}} Normalized\_Cost_i} \right| \quad (5)$$

Here, the normalized cost function $Normalized\_Cost_i$ is defined as equation (6).

$$Normalized\_Cost_i = \cos t_i - \min(\mathbf{COST}) \quad (6)$$

The parameter **COST** is a set of cost values of all the countries and can be represented as equation (7), where $N_{country}$ indicates the number of countries.

$$\mathbf{COST} = \overline{\cos t_1,...,\cos t_i,...\cos t_{N_{Country}}} \quad (7)$$

3) According to the values of the power, the population can be classified into two groups: the colonies and the imperialists. The countries that have higher power are considered as the imperialists; they start to take possession of the colonies which are the countries with the lower power. Thus, the empires are formed.

4) The next step of the sequence is using evolutionary operators. They are applied to update the power of countries by changing the characteristics of the imperialists or colonies. The iterating algorithm converges to the global optimum when there is one empire [23].

The evolutionary operators of ICA can be explained as follows:
1) Assimilation operator: This operator is applied to the weakest colony of each empire and changes the colony's characteristics by assimilating it to their corresponding imperialist. This operator application results in updating the cost function of colonies.

2) Revolution operator: This operator is applied to the imperialist of the weakest empire and causes a change in its characteristics. The goal of the revolution operator is to change some parameters of the individual to prevent the algorithm from falling into local suboptimal solutions

3) Exchange operator: This operator is applied to colonies and imperialists of each empire. For each empire, it compares the power of the imperialist and its colonies and establishes if an imperialist has less power than its corresponding colonies. It updates the position state of colonies by exchanging its position with its corresponding imperialist.

4) Competition operator: This operator is applied to the weakest empire. It picks up the colony with the weakest power and joins it to another empire.

The standard ICA has been applied for solving continuous optimization problems [15] and has shown an excellent convergence characteristic. Thus, it is interesting to test it to resolve combinatorial optimization problems.

### B. ICA-based Channel Allocation

*1) Encoding and Initialization of Popualtion*

In order to apply ICA for channel allocation, we define two terms, province and resource to suggest a new representation. It is referred to as 'grouping imperialist competition algorithm' (GICA) and divides each individual into two parts (see Figure 2).

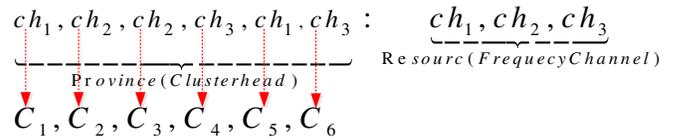

Figure 2. The grouping representation in GICA.

As Figure 2 indicates, each solution can be represented as $\underbrace{ch_1,ch_2,ch_2,ch_3,ch_1}_{\text{Province(Clusterhead)}} : \underbrace{ch_1,ch_2,ch_3}_{\text{Resourc(FrequecyChannel)}}$ and can be encoded by the integer numbers. The number of provinces is equal to the number of clusters and is similar for all of countries. The province part contains information of which channel has been assigned to which clusters; the resource part contains information about the used channels and shows a permutation order of used channels. For each country, the number of resources, $m$, are randomly chosen from a uniform distribution $[0, N_{Avaialble\_channel}]$, where $m$ is the maximum number of available channels. The population for clustered ad hoc networks with $N_{Avaialble\_channel}$ available channels and $N_c$ cluster heads can be represented as Figure 3. The next step is forming empires and choosing the colonies for each empire that is similar to the standard ICA.

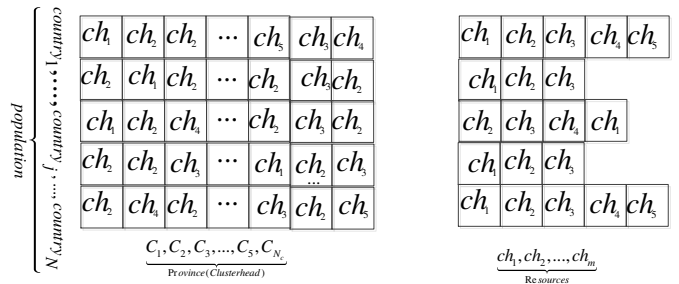

Figure 3. An example of a population in GICA.

*2) Cost function and power:*

We define two cost functions. The first one is a single objective function as (1) and is defined as (8) to find the chromatic number. The second one is a multi-objective function and is defined as equation (9); it is a combination of the optimization function of equation (1) and its constraints as equation (2). For each solution, there is one channel e.g., $ch'$ that has been reused more than others. The number of cluster heads that have been assigned by $ch'$ is represented by $C(ch')$ and is calculated as equation (10).

$$f_1 = \frac{1}{N_{Avaiable\_ch} \times N_c} \sum_{p=1}^{N_{Avaiable\_ch}} \sum_{q=1}^{N_c} x_{pq} \qquad (8)$$

$$f_2 = \frac{1}{N_c \times N_{Avaiable\_ch}} \sum_{p=1}^{N_{Avaiable\_ch}} \sum_{q=1}^{N_c} x_{pq} + \frac{1}{C(ch')} \sum_{p=1}^{N_c} \sum_{q=1}^{N_c} z_{pq} \qquad (9)$$

$$C(ch') = \max_{p=1}^{N_{Avaiable\_channel}} \sum_{q=1}^{N_{cluster}} x_{pq} \qquad (10)$$

*3) Evaluationary Operators*

For the ICA-based channel allocation, the assimilation and revolution are redefined. It should be mentioned that, the evolutionary operators of GICA are applied to the resource part and then the province part is re-arranged according to the resource part. The assimilation operator is applied to the weakest colony of each empire using the following procedures: 1) A randomly selected element of the imperialist's resource is injected to the colony's resource part (it is injected at the first place of the resource part). 2) The province part of the colony is overwritten according to the position of selected elements in the province part of the imperialist. 3) Due to overwriting, some elements of the resource part of the colony might lose their assignment. Thus, they are removed from the resource part. 4) Using the remaining elements in the resource part the province part should be reassigned to the resource part to satisfy the constraints. The steps of the assimilation operator are described by Figure 3.

The revolution operator might have one of the following procedures. Colonies with the lower power are selected and then some elements of their resource parts are removed. Or additional elements are added to the resource part of an imperialist (the imperialist of the weakest empire), After each level, the province part is reassigned according to the changes in the resource part. Due to these operations, the length of each individual is variable.

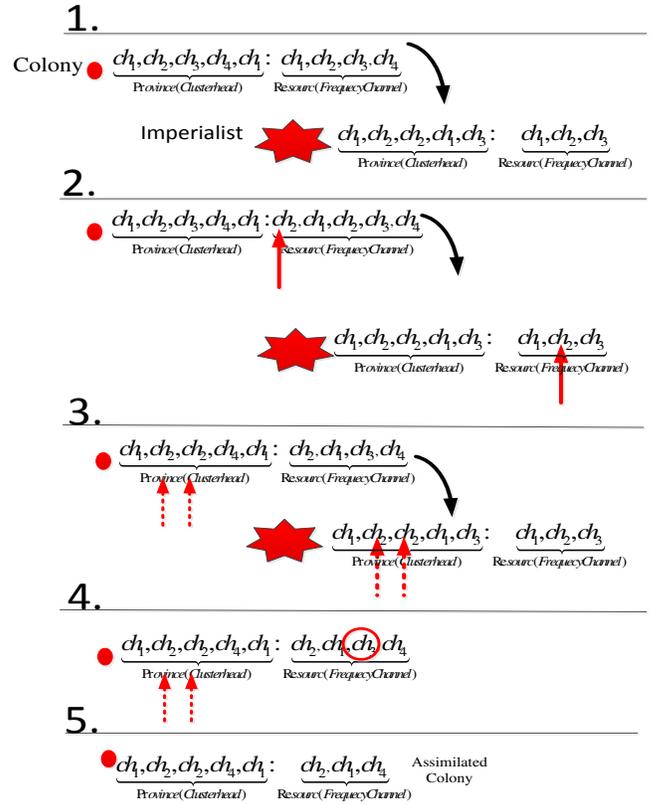

Figure 4. The steps of assimilation operator.

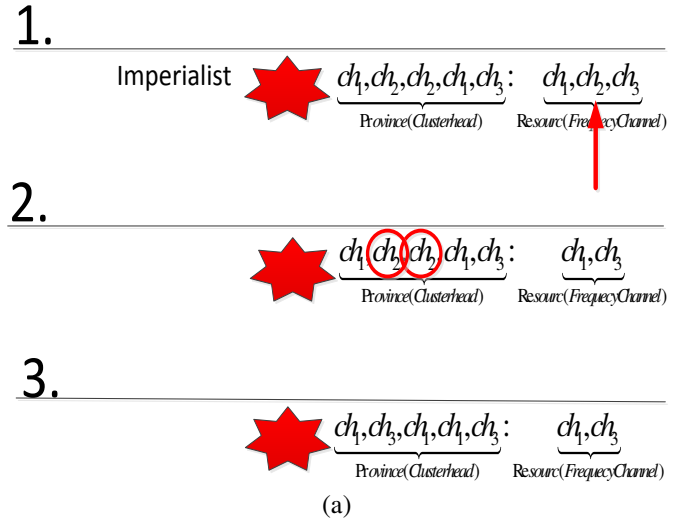

(a)

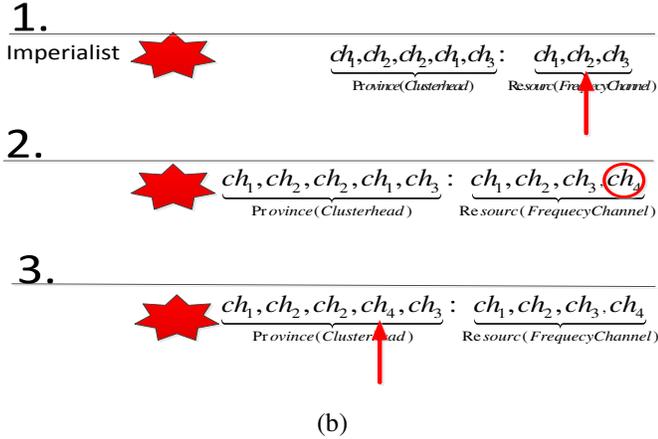

(b)

Figure 5. The steps of revolution operator; (a) demonstrates remove procedure while (b) shows the add procedures.

## V. Simulation

### A. Simulation Model

In this section, we use MATLAB to simulate several scenarios to evaluate the suggested method. A fairly simple model is assumed where $N$ nodes are placed in a 1000 x 1000 meter square. The position of each individual node has two coordinates, $x$ and $y$, each of them is drawn from a uniform distribution [0, 1000]. After generating an ad hoc network, it is clustered by using LID. The main assumptions of the simulated model have previously been described in Section III. It should be noted that the results are compared with the Grouping Genetic Algorithm based method (GGA) that has been explained in [17]. These methods, GICA and GGA, are indexed as ICA and GA in all of figures.

### B. Performance Metrics

The performance of suggested methods is evaluated using two factors: channel reuses efficiency [4] and the fractional interference. These factors are defined as follows:
1. Channel reuse efficiency is determined as the ratio of the number of cluster heads to the number of assigned channels.
2. Fractional interference is defined as the ratio of the number of cluster heads interfering with each other after assigning channels (GICA-based method or GGA-based method) method to the number of interfering cluster heads assigning a single channel to all of the cluster heads.

### C. Simulation Results

At the first experiment, GICA is examined for three different networks with 75 nodes. The networks differ in the node's transmission ranges (TR: 100, 200 and 300 meters). The average number of clusters, available channels, and used channels of single-objective and multi-objective GICA and GGA have been described by the bar chart in Figure 6. It indicates that when the multi-objective function is applied for a large-scale network with more than 30 clusters, the number of used channels by GGA is smaller than GICA. It is also important to note that in networks with an average of 5 and 10 clusters, the number of used channels have no significant differences when examining multi-objective or single-objective optimization functions. Figure 7 depicts the values of channel reuse efficiency for different methods (i.e., single-objective and multi-objective of GGA and GICA). It shows that the high spectral efficiency is achieved by using the ICA-based method for the single objective function. However, the channel allocation scheme is not feasible from the perspective of co-channel interference. As Figure 8 indicates, the value of fractional interference from applying GGA and GICA for a single objective is noticeably higher and can substantially degrade the communication performance. While for multi-objective function, the fractional interference of both GGA and GICA are equal with zero. Figure 8 indicates that GICA has better results in terms of fractional interference.

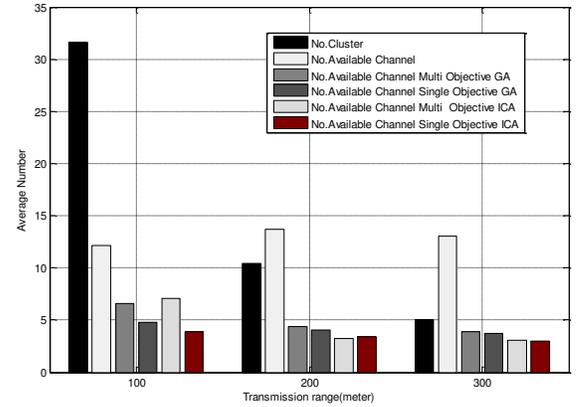

Figure 6. Bar charts of the average number of clusters, available channels and used channels. The values have been obtained by applying GICA and GGA to optimize the single-objective function and multi-objective function.

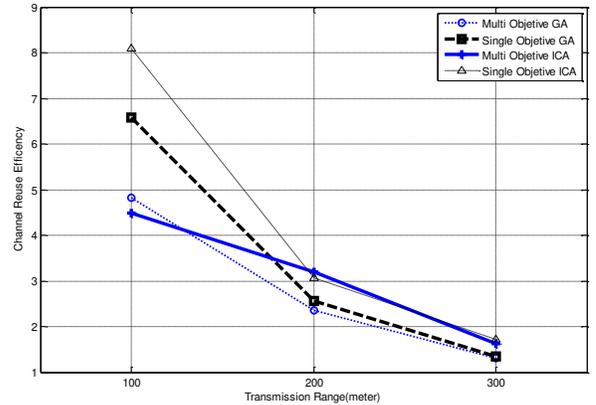

Figure 7. The obtained values of channel reuse efficiency using GICA and GGA. For multi-objective function, the channel reuse efficiency is lower than the single-objective function (blue solid and dotted lines).

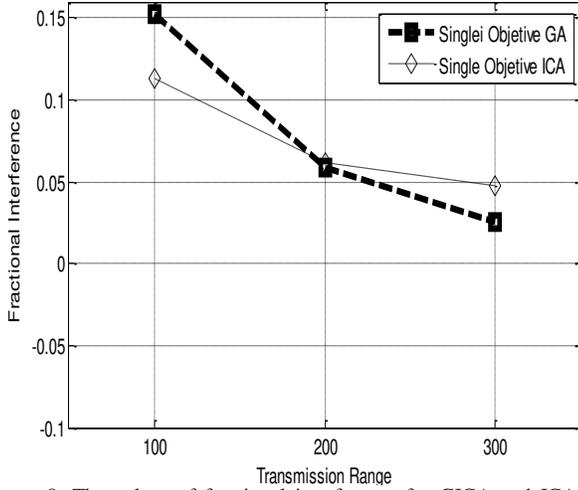

Figure 8. The values of fractional interference for GICA and ICA channel assignment schemes. For a network with a large number of clusters, approx. 30 clusters, ICA has lower fractional interference (see dashed-dotted line for transmission range of 100).

TABLE I. COMPARISON BETWEEN THREECHANNEL ALLOCATION METHODS FOR MANETS WITH DIFFERENT SIZES.

| Method | Specification | | |
|---|---|---|---|
| | No. of Nodes | Channel reuse efficiency | No. of Assigned Channels |
| GICA | 100 | 1.8 | 2.95 |
| | 200 | 1.85 | 2.8 |
| | 300 | 1.78 | 2.85 |
| GICA(two) | 100 | 1.7 | 3.1 |
| | 200 | 1.7 | 3.5 |
| | 300 | 1.728 | 2.95 |
| Channel_Segration [4] | 100 | 1.7000 | 4.1 |
| | 200 | 1.7000 | 5.7 |
| | 300 | 3.1000 | 9 |
| Greedy-based [4] | 100 | 4.7000 | 11 |
| | 200 | 3.7000 | 13 |
| | 300 | 3.2000 | 15 |

As the second experiment, GICA is applied for three different scenarios with 100, 200 and 300 nodes. In all cases, the transmission range is fixed and is set to 300 meters. Table I compares the results of GICA with the results of a greedy based method and another method known as the channel segregation method. For a large size network, the channel segregation method has the highest value of channel reuse efficiency.

While for medium size networks, GICA performs better than the channel segregation method in terms of channel reuse efficiency. It should be noted that the channel reuse efficiency is a factor that is related to the number of clusters. Thus, for the network with 300 nodes, in spite of a large number of used channels (i.e., 9), its channel reuse efficiency is higher than GICA.

As a final experiment, the convergence characteristics of GICA and GGA are investigated. The GICA is applied to find a near optimal spectrum allocation to a network with 300 nodes and 30 clusters. In the case of GICA, the number of countries is chosen as 50 and 6 countries form the empires. In GGA, the number of chromosomes to create the initial population is 50. Figure 9 shows the curve of interference power, the number of used channels and the objective function, versus number of iterations. It can be observed that that using a single-objective optimization, the number of used channels are smaller than for the multi-objective methods (see Figure 9.(a) that shows the number of used channels versus number of iterations). It is also noticeable that for the single-objective function, the number of used channels by ICA is smaller than for GGA. In this case, after channel allocation to the network, the average of co-channel interference between the clusters obtained by GICA is equal to -13 dBm, which is smaller than the counterpart values from GGA (see Figure 9.(b)). The multi-objective GICA cannot provide as high spectral efficiency as multi-objective GGA can (see Figure 9.(a)). However, the multi-objective GICA induces smaller interference in the network than multi-objective GGA does (see Figure 9.(b)). The average co-channel interference power that is obtained from using a single channel to all nodes is equal to -9 dBm. Thus, an ICA-based channel allocation scheme can minimize the co-channel interference power in the network.

The values of objective functions versus iterations is displayed in the graph in Figure 9.(c). The fast convergence characteristic of GICA, in comparison to GGA, is obvious from the graphs in Figure 9.(c). The multi-objective GICA converges after 60 iterations, while the multi-objective GGA converges after 160 iterations. For the case using a single objective function, GICA also performs better than GGA to find a near optimal solution. It can be observed that the dashed-dotted line converges to zero before 20 iteration; while the dotted line (is related to single-objective GGA) does not converge to zeros even after 200 iterations.

The effects of exploitation rate in GGA and GICA have been investigated through changing the mutation rate and revolution rate. As Figure 10 indicates, using a higher exploitation rate the evolutionary algorithms (GGA and GICA) converge quicker. However, the speed of convergence in GICA is higher than GGA.

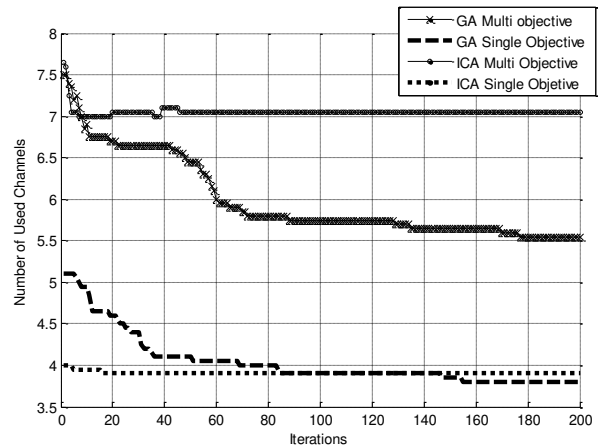

(a)

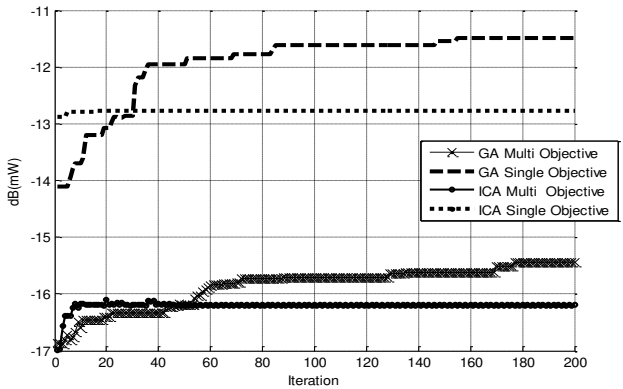

(b)

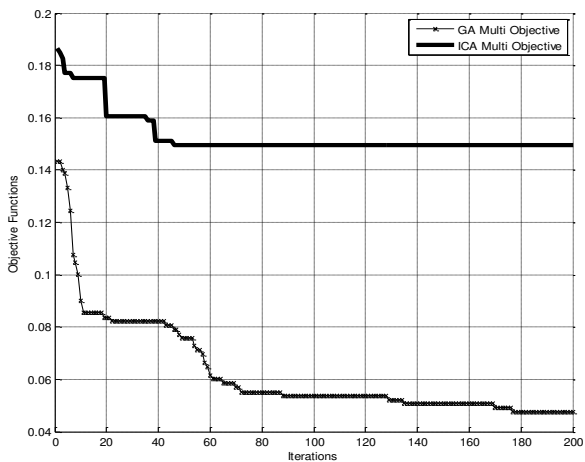

(c)

Fig. 9. Demonstrates the convergence behavior of GGA and GICA for a network with 30 cluster heads. (a). The number of used channels. (b) The average of interference power. (c). The values of objective functions during the iterations.

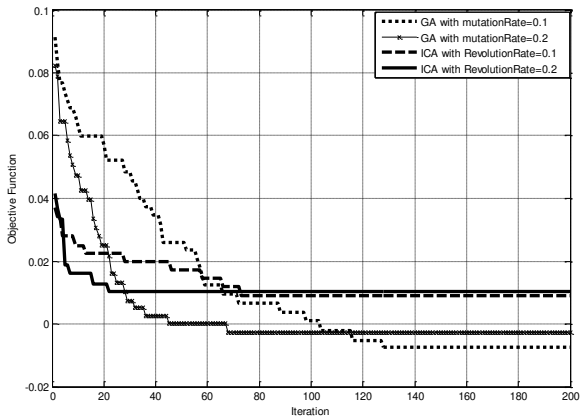

Fig. 10. The convergence behavior of GGA and GICA using diferent parameters.

## VI. CONCLUSION

The main contribution of this paper is to introduce a suitable meta-heuristic method and design a feasible objective function for the spectrum allocation problem in a clustered wireless ad hoc network topology. This proposed method is a meta-heuristic method and it is referred to as ICA [16]. We presented a multi-objective function with the capability to make a tradeoff between interference power and the number of used channels. On the other hand, the proposed method assigns available channels to the clusters, with a high spectral efficiency avoiding co-channel interference. The suggested method is evaluated by several simulation experiments for some scenarios in terms of fractional interference and channel reuse efficiency. The results are also compared with the performance of GGA. The obtained results verify that the GICA has the capability to approximate the Pareto solutions to minimize the average level of interference power (see Figure 9. (b) ) and maximize spectral efficiency (see figure 9. (c)). The conducted simulation experiments also indicate that GICA has the capability to converge very quickly (see Figure 9. (a) and (c)).

As future works, a clustering scheme on the basis of ICA will be used instead of the present lowest ID clustering algorithm. The network model will include the mobility and traffic patterns and ICA will be investigated for other optimization problems (e.g., power control that limits the increase in interference power when the number of nodes increases). The distributed versions of ICA will also be developed and channel allocation for both inter-cluster and intra-cluster communication will be considered.